\documentclass[aps,prl,reprint,superscriptaddress,nofootinbib,nobibnotes,floatfix]{revtex4-2}

\usepackage{amsmath,amssymb,bm}
\usepackage{graphicx}
\usepackage{hyperref}
\hypersetup{hidelinks}

\begin{document}

\title{Current-Driven Magnetization Switching via Zhang-Li Torque}

\author{Shuai Li\textsuperscript{1}, Maokang Shen\textsuperscript{2,\ensuremath{\dagger}}, Liuyu Tang\textsuperscript{2}, Weiwei Lin\textsuperscript{3}, Kaiming Cai\textsuperscript{4}, Tao Wang\textsuperscript{1}, Tianli Jin\textsuperscript{1}, and Yue Zhang\textsuperscript{1,\ensuremath{\ddagger}}\\[0.4em]
{\small\itshape \textsuperscript{1}School of Integrated Circuit, Huazhong University of Science and Technology, Wuhan, China\\
\textsuperscript{2}School of Integrated Circuits, Hubei University, Wuhan, China\\
\textsuperscript{3}School of Physics, Southeast University, Nanjing, China\\
\textsuperscript{4}School of Physics, Huazhong University of Science and Technology, Wuhan, China}}
\noaffiliation

\begin{abstract}
Current-driven magnetization switching is currently governed by two established mechanisms: the Slonczewski spin-transfer torque (STT) and the spin-orbit torque (SOT). Here we identify the Zhang--Li STT as a third mechanism that operates on a different physical footing. Using micromagnetic simulations, we show that the interfacial Dzyaloshinskii--Moriya interaction (iDMI) inevitably creates spatial magnetization gradients at device edges, enabling the Zhang--Li torque to drive deterministic, field-assisted switching that fully reproduces all hallmark behaviors of SOT, including polarity control and effective-field characteristics. Our work establishes the Zhang--Li torque as an autonomous switching pathway, redefining the physical picture of current-induced magnetization reversal and providing new design principles for spintronic devices.
\end{abstract}

\maketitle

\begingroup
\renewcommand{\thefootnote}{}
\footnotetext{\textsuperscript{\ensuremath{\dagger}} Corresponding author: maokangshen@hubu.edu.cn\\
\textsuperscript{\ensuremath{\ddagger}} Corresponding author: yue-zhang@hust.edu.cn}
\endgroup

Current-driven magnetization switching is the physical foundation of modern nonvolatile spintronic memory and computing technologies \cite{r1,r2}. To date, two principal mechanisms have been established to govern such switching: the Slonczewski STT, which relies on a spin-polarized current generated from a reference layer \cite{r3,r4,r5}, and the SOT originating from spin currents in a nonmagnet (NM)/magnet (M) heterostructure \cite{r6,r7,r8,r9,r10,r11,r12,r13,r14}. Between the two, SOT has attracted particular attention due to its superior faster switching speed, establishing itself as the leading candidate for next-generation magnetic random-access memory (MRAM) with fast processing. To date, extensive studies have established that pure spin currents in SOT devices---originating from the spin Hall effect in heavy metals \cite{r6,r7,r8}, the Rashba effect at NM/M interfaces \cite{r9,r10,r11}, spin--momentum locking in topological insulators (Tl) \cite{r12,r13}, or chiral spin currents in Weyl semimetals \cite{r14}---constitute the main driving force.

Interestingly, current-induced switching reminiscent of SOT has also been observed in magnetic systems that themselves serve as the primary conduction channel, including composition-graded alloys \cite{r15,r16}, magnetic semiconductors \cite{r17}, and surface-oxidized Ni films \cite{r18}. While this observation naturally invokes STT, the Slonczewski formalism cannot account for it, as it necessitates a spin-polarized current injected with a noncollinear component relative to the local magnetization---a condition absent for a simple charge current injected in a magnet.

Beyond the conventional Slonczewski STT, the Zhang--Li torque represents a distinct mechanism inherently capable of driving the dynamics of nonuniform magnetic textures, such as domain walls (DWs) and skyrmions \cite{r19,r20,r21}. In magnetic thin films where inversion symmetry is broken along the thickness direction, the presence of such nonuniformities is fundamentally mandated by the iDMI \cite{r22,r23}. Specifically, the iDMI enforces twisted boundary conditions that inevitably tilt the magnetization at device edges, thereby giving rise to DW-like structures \cite{r24,r25} (Figure 1), which can be quantified by Eq.~\eqref{eq:boundary} \cite{r25}:

\begin{equation}
  \frac{d\mathbf{m}}{dn}=\frac{D}{2A}(\hat{\mathbf{e}}_z\times\hat{\mathbf{n}})\times\mathbf{m}.
  \label{eq:boundary}
\end{equation}

Here A, \ensuremath{\hat{\mathbf{e}}_z}, and \ensuremath{\hat{\mathbf{n}}} are the exchange stiffness constant, the unit vectors pointing from the NM layer to the M layer and that for the normal direction at the device edge, respectively. Consequently, the switching process does not proceed via coherent magnetization rotation, as presumed in early macrospin models, but instead involves edge nucleation followed by DW propagation \cite{r26}.

The ubiquitous boundary textures induced by the iDMI provide a natural and unavoidable conduit through which the Zhang--Li torque can act. This texture-mediated switching scenario immediately raises a critical question: beyond the extensively studied spin--orbit torque (SOT), could the Zhang--Li STT play a substantial role in the switching dynamics? In principle, this torque could modify the boundary DW configuration and thereby influence the overall switching behavior. Yet, this intriguing possibility has remained entirely unexplored.

In this work, we employ micromagnetic simulations to systematically investigate the role of the Zhang--Li torque in iDMI-enabled switching within a ferromagnetic (FM) layer. Our results reveal that this torque can reproduce the benchmark switching characteristics typically associated with SOT, including the necessity of a current-collinear magnetic field, the controllability of reversal polarity, and the emergence of an effective magnetic field. These findings demonstrate that the Zhang--Li torque constitutes an essential driving force for magnetic switching in structurally asymmetric FM layers. Overall, this work offers a fresh perspective on current-driven switching dynamics and provides valuable insights for the design and optimization of spintronic devices.

\begin{figure}[!htbp]
  \centering
  \includegraphics[width=0.78\columnwidth]{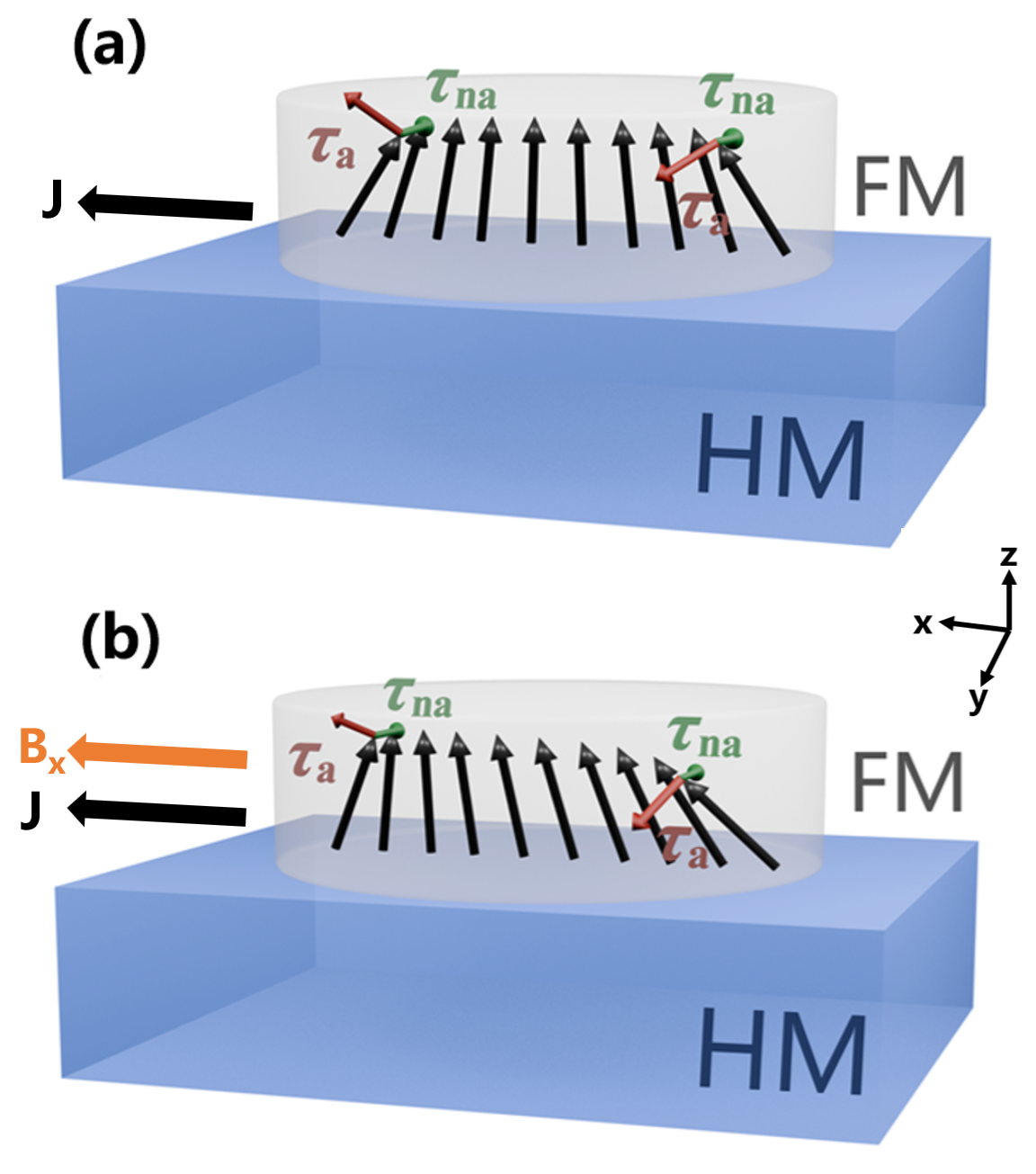}
  \caption{Device geometry and torque configuration. (a) Perpendicularly magnetized ferromagnetic (FM) nanodisk on a heavy-metal (HM) layer, with charge current J applied along the x direction. In the absence of a longitudinal magnetic field B\textsubscript{x}, the iDMI cants the edge moments, giving rise to symmetric adiabatic (\ensuremath{\tau}\textsubscript{a}) and nonadiabatic (\ensuremath{\tau}\textsubscript{na}) Zhang-Li torques under the current drive. (b) Asymmetric torque configuration induced by current J in the presence of a longitudinal field B\textsubscript{x}.}
  \label{fig:1}
\end{figure}

Our calculations are based on the Landau--Lifshitz--Gilbert (LLG) equation incorporating the Zhang--Li torque, expressed as
\begin{equation}
  \boldsymbol{\tau}_{\mathrm{ZL}}=-b_J\mathbf{m}\times\left(\mathbf{m}\times\frac{\partial\mathbf{m}}{\partial x}\right)-c_J\left(\mathbf{m}\times\frac{\partial\mathbf{m}}{\partial x}\right),
  \label{eq:zl-total}
\end{equation}
The torque in Eq.~\eqref{eq:zl-total} contains an adiabatic term and a nonadiabatic term \cite{r19,r27}, respectively given by Eqs.~\eqref{eq:tau-a} and \eqref{eq:tau-na}:
\begin{align}
  \boldsymbol{\tau}_{a\phantom{n}} &= -b_J\mathbf{m}\times\left(\mathbf{m}\times\frac{\partial\mathbf{m}}{\partial x}\right),
  \label{eq:tau-a}\\
  \boldsymbol{\tau}_{na} &= -c_J\left(\mathbf{m}\times\frac{\partial\mathbf{m}}{\partial x}\right).
  \label{eq:tau-na}
\end{align}
Here
\begin{align}
  b_J &= \frac{JPg\mu_{\mathrm B}}{2eM_s(1+\beta^2)}\approx\frac{JPg\mu_{\mathrm B}}{2eM_s},
  \label{eq:b-j}\\
  c_J &= b_J\beta.
  \label{eq:c-j}
\end{align}
The coefficients in Eqs.~\eqref{eq:b-j} and \eqref{eq:c-j} contain $J$, $g$, $\mu_{\mathrm B}$, $e$, and $M_s$, which denote the current density, Land\'e factor of an electron, Bohr magneton, elementary charge magnitude, and saturation magnetization, respectively. The parameter $\beta$ is the nonadiabatic coefficient and is usually much smaller than 1, while $P$ is the spin polarization in the magnetic layer. The quantity $\partial\mathbf{m}/\partial x$ denotes the spatial gradient of the magnetization unit vector along the direction of the electrical current density, i.e., the $x$-axis direction in this work (Fig. 1).

Fig. 1 illustrates the essential physical picture. In the absence of the magnetic field, the iDMI renders the magnetization orientations at the left and right domain boundaries symmetric about the central axis. Under a fixed current density J, the Zhang-Li torques acting on the magnetic moments of the left and right boundaries are equal in magnitude but differ in direction. Specifically, the \ensuremath{\tau}\textsubscript{na} components are identical in magnitude and both align along the y direction, whereas the \ensuremath{\tau}\textsubscript{a} components have identical x-components but opposite z-components [Fig. 1(a)]. Consequently, without the assistance of an external magnetic field, the Zhang-Li torque alone cannot induce magnetization switching.

When a longitudinal magnetic field B\textsubscript{x} is applied collinearly with the current direction, it breaks the symmetry of the magnetization orientations at the two lateral boundaries of the nanodisk. Taking the configuration in Fig. 1(b) as an example, the downward z-component of the Zhang-Li torque at the right edge becomes larger in magnitude than the upward counterpart at the left edge, yielding a net z-component of \ensuremath{\tau}\textsubscript{a} (\ensuremath{\tau}\textsubscript{az}) that enables deterministic switching. From Eqs.~\eqref{eq:zl-total}--\eqref{eq:tau-na}, it readily follows that the sign of \ensuremath{\tau}\textsubscript{az} changes and such switching does not occur upon reversing the sign of J, B\textsubscript{x}, or D, individually, unless the magnetization itself is reversed. This switching behavior is consistent with the conventional magnetic-field-assisted SOT switching paradigm.

To verify the above scenario, we perform COMSOL and micromagnetic simulations, which includes the Zhang-Li torque as a built-in STT module \cite{r28}. (See more details in the Supplementary Materials Secs. S1 and S2). The simulated system adopts the standard SOT-MRAM pillar geometry: a perpendicularly magnetized nanodisk with radii ranging from 40 to 250 nm, grown on a heavy-metal underlayer. The material parameters are chosen to represent typical HM/FM bilayers with strong iDMI, such as Pt/Co \cite{r29}. To isolate the intrinsic contribution of the Zhang-Li torque to switching, most simulations are performed under the assumption of temperature T = 0 K. We also examine thermal effects on switching at 300 K \cite{r30}. Although the Co layer is usually much thinner than Pt, the current density in Co is very close to that in Pt (Figure S2 in the Supplementary Materials). This confirms the nonnegligible role of the current flowing within the Co layer, a feature that has been widely corroborated by experiments \cite{r31,r32,r33,r34}.

\begin{figure}[!htbp]
  \centering
  \includegraphics[width=\columnwidth]{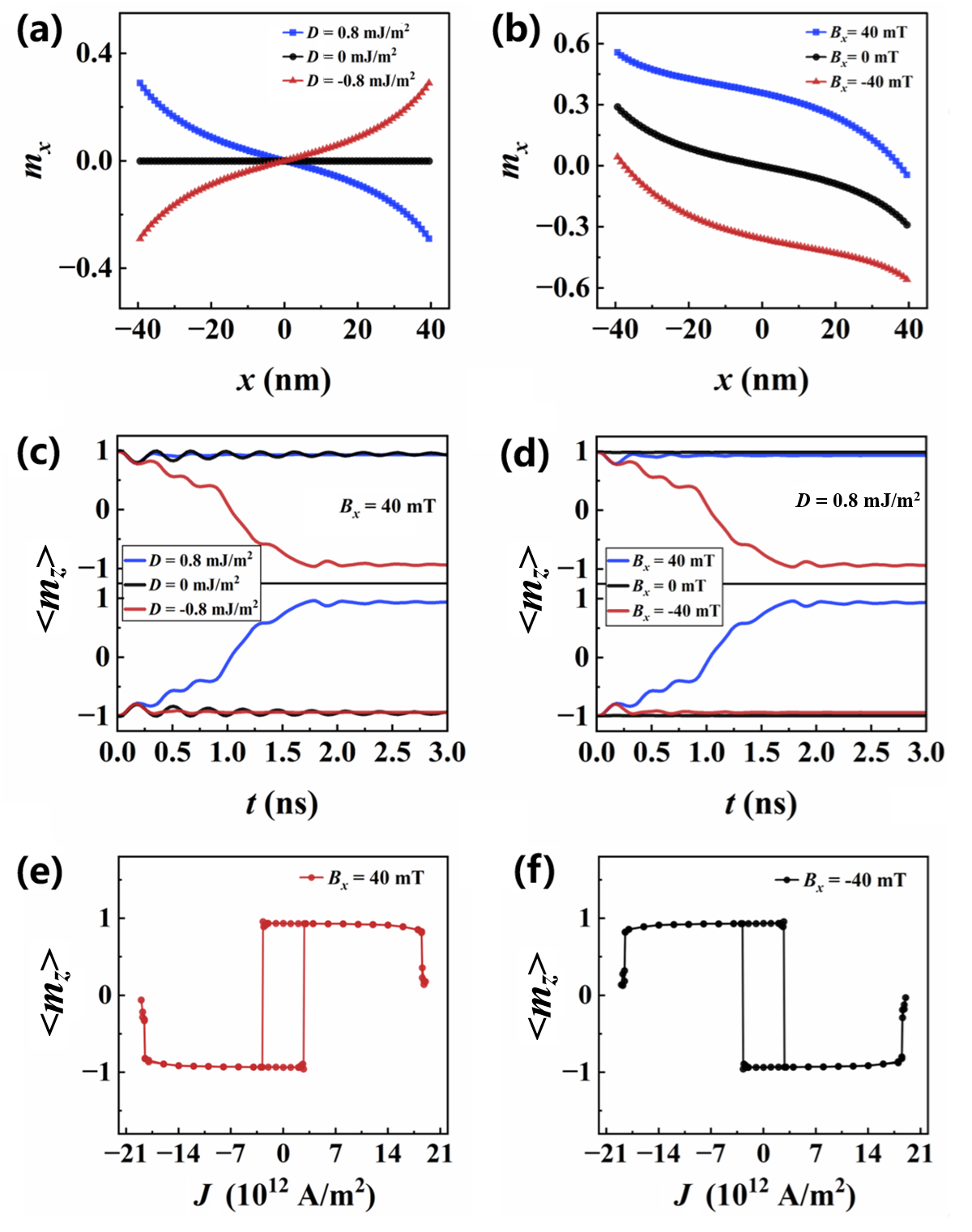}
  \caption{Control and dynamics of iDMI-assisted Zhang-Li-torque switching. (a), (b) Distribution of m\textsubscript{x} along the diameter in the x-axis direction. (c), (d) Time evolution of the spatially averaged m\textsubscript{z} (\ensuremath{\langle m_z\rangle} ) (e), (f) \ensuremath{\langle m_z\rangle}-J switching loops for B\textsubscript{x}= +40 and -40 mT.}
  \label{fig:2}
\end{figure}

Figures 2 (a) and (b) show the magnetization configuration of the nanodisk with different D. For D = 0 mJ/m\textsuperscript{2}, in which the magnetization is uniformly aligned along the x direction. For D = 0.8 mJ/m\textsuperscript{2} and D = -0.8 mJ/m\textsuperscript{2}, the edge magnetization is tilted, while the overall magnetization configuration remains symmetric about the center of the disk (see Supplemental Material, Sec. S3, for more details). The combined action of iDMI and an external field B\textsubscript{x} produces pronounced asymmetric canting at the disk boundary.

Figures 2 (c) and (d) present the time evolution of the spatially averaged out-of-plane magnetization \ensuremath{\langle m_z\rangle} under various parameter combinations, with \ensuremath{\alpha} = \ensuremath{\beta} = 0.03 and a fixed current density J = 4 \ensuremath{\times} 10\textsuperscript{12} A/m\textsuperscript{2}. For D = \ensuremath{\pm} 0.8 mJ/m\textsuperscript{2} and \ensuremath{B_x} = + 40 mT, the system exhibits complete switching from both the initial up (\ensuremath{\langle m_z\rangle} = +1) and down (\ensuremath{\langle m_z\rangle} = -1) states within approximately 2 ns. The switching polarity is strictly determined by the sign of D: for D = +0.8 mJ/m\textsuperscript{2}, \ensuremath{\langle m_z\rangle} switches from +1 to -1, while for D = -0.8 mJ/m\textsuperscript{2}, it switches from -1 to +1, under identical current and field conditions. This sign dependence directly reflects the chiral nature of the iDMI-induced edge magnetization: opposite iDMI chirality produces opposite edge canting, which in turn reverses \ensuremath{\tau}\textsubscript{az}. In contrast, for D = 0 mJ/m\textsuperscript{2}, no switching occurs regardless of \ensuremath{B_x} magnitude. Without edge canting in the absence of iDMI, the torque vanishes because the \ensuremath{\frac{\partial\mathbf{m}}{\partial x}} is zero in a uniformly magnetized state.

In NM/M heterostructures, the sign of D can be reversed by varying the composition of the NM layer, e.g., using Pt versus W. Alternatively, in NM(\ensuremath{t_1})/M/NM(\ensuremath{t_2}) sandwich structures with identical NM composition but different layer thicknesses \ensuremath{t_1} and \ensuremath{t_2}, the iDMI sign can also be flipped by swapping the order of the bottom and top NM layers. Experimentally, in heterostructures with opposite D, the switching polarity is consistently found to be reversed \cite{r35}. For D = +0.8 mJ/m\textsuperscript{2}, switching occurs for both \ensuremath{B_x} = +40 mT and \ensuremath{B_x} = -40 mT, but with opposite final states: \ensuremath{B_x} = +40 mT drives the switching of magnetization from up to down, while \ensuremath{B_x} = -40 mT drives it from down to up. At \ensuremath{B_x} = 0 mT, the magnetization oscillates around its initial state. Figures 2(a) and 2(b) confirm that the iDMI-generated edge DWs are strictly necessary for the Zhang--Li torque to exert a net symmetry-breaking effect under the assistance of B\textsubscript{x}.

Figures 2(e) and 2(f) display the \ensuremath{\langle m_z\rangle} versus J hysteresis loops for \ensuremath{B_x} = +40 mT and -40 mT, respectively. For \ensuremath{B_x} = +40 mT, the loop is counterclockwise: increasing \ensuremath{J_x} from zero drives \ensuremath{\langle m_z\rangle} from +1 to -1 at a critical current density J\textsubscript{c} = +2.78 \ensuremath{\times} 10\textsuperscript{12} A/m\textsuperscript{2}, while reversing \ensuremath{J_x} brings it back at J\textsubscript{c} = -2.78 \ensuremath{\times} 10\textsuperscript{12} A/m\textsuperscript{2}. For \ensuremath{B_x} = -40 mT, the loop is clockwise, with the switching branches reversed. At higher current densities (J \ensuremath{>} 1.8 \ensuremath{\times} 10\textsuperscript{13} A/m\textsuperscript{2}), the in-plane component of the Zhang--Li torque gradually reduces \ensuremath{\langle m_z\rangle} to near zero under strong in-plane components of the torque. These loop features---field-polarity-dependent branch selection, symmetric critical currents, and the decreased \ensuremath{\langle m_z\rangle} at a high current density---are hallmarks of SOT switching reported in experiments \cite{r7,r36,r37}, demonstrating that the Zhang--Li torque can replicate the macroscopic phenomenology without invoking SOT.

\begin{figure}[!htbp]
  \centering
  \includegraphics[width=\columnwidth]{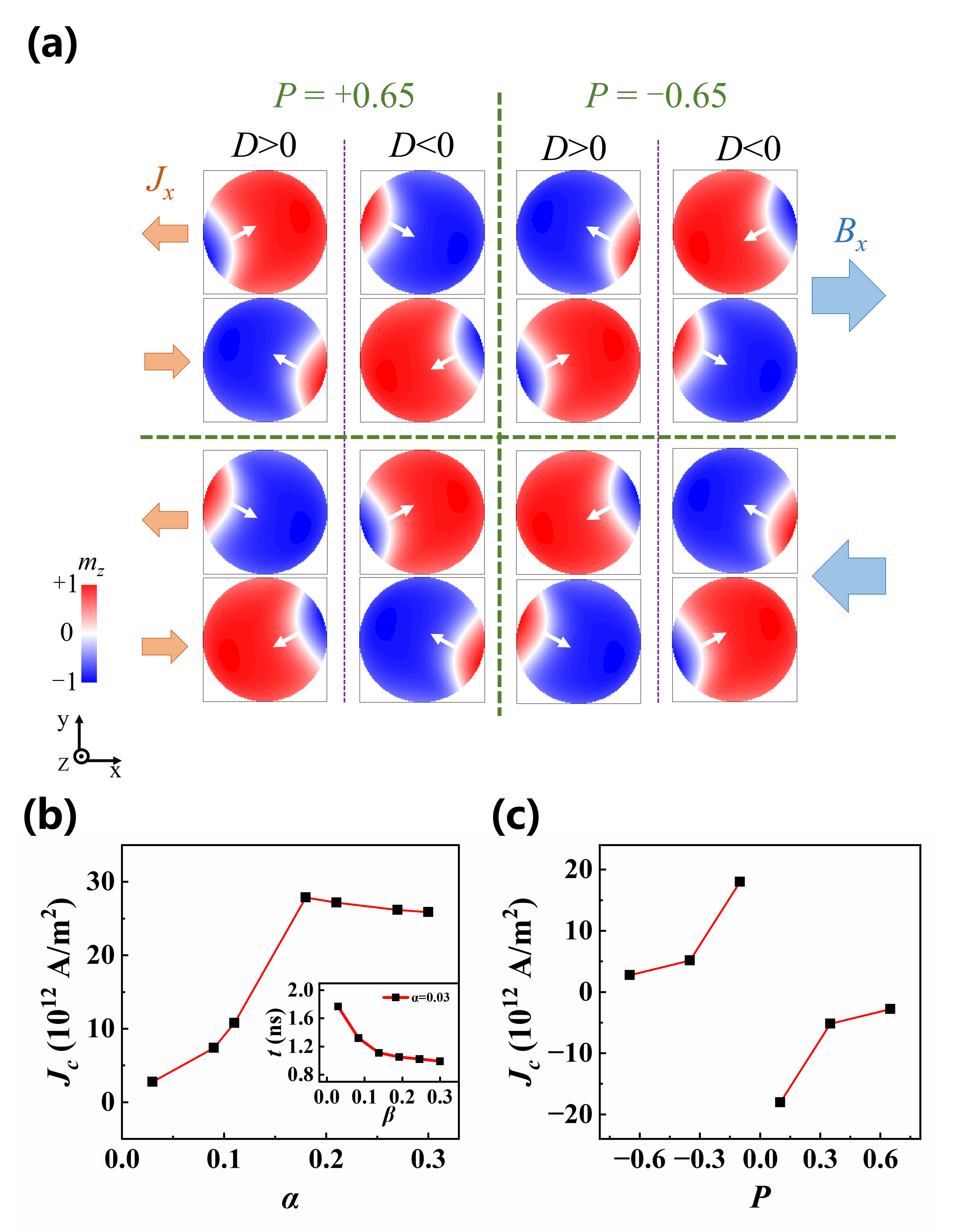}
  \caption{Symmetry tests of iDMI-assisted switching. (a) Time-resolved m\textsubscript{z} maps for the sign combinations of D, B\textsubscript{x}, and J at P = +0.65 and -0.65, respectively. (b) and (c) Influence of Gilbert damping coefficient, nonadiabatic factor, and P on the critical current density (J\textsubscript{c})}
  \label{fig:3}
\end{figure}

Figure 3 provides a detailed analysis of the spatiotemporal dynamics of switching. Figures 3(a) show time-resolved spatial maps of m\textsubscript{z} for all sign combinations of D, \ensuremath{B_x}, and \ensuremath{J_x}, at P = +0.65 and -0.65, respectively. In all cases where switching occurs, the process consistently starts with the nucleation of a reversed domain at one edge of the disk (either left or right), followed by DW propagation across the disk to the opposite edge. We also include the Oersted field arising from the applied current itself, and the Oersted field modifies the nucleation site between up and down but does not influence the polarity of switching (see the Supplemental Material S4).

For a fixed set of D, \ensuremath{B_x}, and \ensuremath{J_x}, reversing P changes the switching polarization and reverses the direction of DW motion. For example, with P \ensuremath{>} 0 and \ensuremath{J_x} \ensuremath{>} 0, the DW nucleates at the right edge and moves leftward (i.e., opposite to the current, along the electron flow). With P \ensuremath{<} 0, under otherwise identical conditions, the DW nucleates at the left edge and moves rightward (along the current). This reversal is a direct consequence of the sign change in the torque magnitude: the Zhang--Li torque is proportional to P, so a negative P reverses the force on the DW, flipping its propagation direction. On the other hand, for a fixed sign of P, changing the sign of D also reverses the polarization of switching due to the changing of chirality for magnetization structure in the boundary DW.

The Gilbert damping coefficient \ensuremath{\alpha} and the nonadiabatic factor \ensuremath{\beta} are two key parameters in the Zhang--Li torque model. In the simulations described above, with \ensuremath{\alpha} = \ensuremath{\beta} = 0.03, the switching time is about 2 ns. We then vary \ensuremath{\alpha} from 0.03 to 0.3 while keeping the ratio \ensuremath{\beta}/\ensuremath{\alpha} = 1 fixed. At 0 K, the critical current density J\textsubscript{c} increases from 2 \ensuremath{\times} 10\textsuperscript{12} A/m\textsuperscript{2} with \ensuremath{\alpha} for 0.03 \ensuremath{\le} \ensuremath{\alpha} \ensuremath{\le} 0.1 and saturates at approximately 2 \ensuremath{\times} 10\textsuperscript{13} A/m\textsuperscript{2} for \ensuremath{\alpha} \ensuremath{>} 0.1. Given that \ensuremath{\beta}/\ensuremath{\alpha} is likely greater than unity \cite{r38,r39}, we further increase \ensuremath{\beta} while holding \ensuremath{\alpha} = 0.03. Varying \ensuremath{\beta}/\ensuremath{\alpha} does not affect J\textsubscript{c}, but reduces the switching time to below 1 ns, which is comparable to the typical switching speed of SOT-MRAM [Figs. 2(e) and 2(f)] \cite{r18}. We also examine the dependence of J\textsubscript{c} on the spin polarization P over the range from -0.65 to 0.65. The critical current density increases as \ensuremath{|P|} decreases, reaching about 2 \ensuremath{\times} 10\textsuperscript{13} A/m\textsuperscript{2} when \ensuremath{|P|} = 0.1; reversing the sign of P reverses the sign of J\textsubscript{c}. Similar modifications of J\textsubscript{c} are achieved by varying D, B\textsubscript{x}, K, and R within experimentally feasible ranges (see Supplemental Material, Sec. S5).

This observation in Fig. 3 is particularly significant when interpreted in the context of experimental reports on Pt/Co systems. In the conventional SOT picture, the spin current arising from the spin Hall effect in a Pt layer (e.g., in Pt/Co/AlO\textsubscript{x}) provides a fixed spin polarization direction for a given current direction. Since the sign of D in such structures is also fixed, the switching polarity can only be reversed by reversing the current or the in-plane field B\textsubscript{x}; correspondingly, the DW is expected to always move along the current direction. However, reported experiments have shown that in Pt/Co systems with fixed Pt composition, DWs can move either along or opposite to the electron flow \cite{r31,r33,r34,r40}, indicating that SOT alone is insufficient to fully account for the observed switching behavior.

Notably, a growing body of theoretical and experimental work---including giant magnetoresistance measurements---has revealed that the spin polarization P of Co in Pt/Co systems itself can be either negative or positive, depending on the Co thickness and the local magnetic texture \cite{r31,r32,r41}. This provides a natural route for the Zhang--Li STT to come into play. Under negative P, the Zhang--Li torque alone can drive the DW opposite to the electron flow (the so-called negative STT), as illustrated in Fig. 3(a). Importantly, when using the iDMI constant appropriate for Pt/Co, the simulated switching characteristics under negative P are consistent with experimental observations \cite{r18}. Thus, the Zhang--Li torque either override or compete with the SOT contribution in switching. It is also worth noting that such current-driven DW motion against the electron flow is not restricted to Pt/Co. Very recently, similar behavior has been observed in NiCo\textsubscript{2}O\textsubscript{4} without any additional NM layer \cite{r42}, which is attributed to negative P arising from the dominant charge carriers with minority spin polarization at the Fermi level \cite{r43}. This further suggests that the direction of DW motion relative to the current cannot be taken as a smoking-gun signature for distinguishing SOT from STT.

In Pt(t\textsubscript{1})/Co/Pt(t\textsubscript{2}) multilayers, the SOT mechanism predicts that the signs of both D and the total spin polarization can be reversed by interchanging t\textsubscript{1} and t\textsubscript{2}. Hence, besides reversing the current or B\textsubscript{x}, the switching polarity can also be reversed by reversing the Pt layer sequence. In contrast, for the Zhang--Li STT, such a sequence reversal does not change the sign of P. Nevertheless, it can still reverse the switching polarity through a different channel---namely, by altering the sign of D and thus the chirality of the boundary DW, as shown in Fig. 3(a). This distinction further highlights the fundamentally different roles of SOT and STT in determining the switching behavior.

Based on the LLG equation incorporating the Zhang--Li STT, the effective fields corresponding to the adiabatic and nonadiabatic contributions of the Zhang--Li torque can be expressed as Eqs.~\eqref{eq:h-a} and \eqref{eq:h-na}:
\begin{alignat}{2}
  &(\mathbf{H}_{\mathrm{ZL}})_a &&=\frac{b_J}{\gamma}\left(\mathbf{m}\times\frac{\partial\mathbf{m}}{\partial x}\right),
  \label{eq:h-a}\\
  &(\mathbf{H}_{\mathrm{ZL}})_{na} &&=\frac{c_J}{\gamma}\frac{\partial\mathbf{m}}{\partial x}.
  \label{eq:h-na}
\end{alignat}
The $x$- and $y$-components of the total effective field $H_{\mathrm{ZL},x}$ and $H_{\mathrm{ZL},y}$ are quantified by evaluating the spatial distributions of $m_x$, $m_y$, and $m_z$ and their derivatives with respect to $x$ under current injection, as given by Eqs.~\eqref{eq:h-x} and \eqref{eq:h-y}:

\begin{align}
H_{\mathrm{ZL},x} &= \frac{b_J}{\gamma}\left(m_y\frac{\partial m_z}{\partial x}-m_z\frac{\partial m_y}{\partial x}\right)+\frac{c_J}{\gamma}\frac{\partial m_x}{\partial x},
\label{eq:h-x}\\
H_{\mathrm{ZL},y} &= \frac{b_J}{\gamma}\left(m_z\frac{\partial m_x}{\partial x}-m_x\frac{\partial m_z}{\partial x}\right)+\frac{c_J}{\gamma}\frac{\partial m_y}{\partial x}.
\label{eq:h-y}
\end{align}

As shown in Figs. 4(a)-4(d), the adiabatic components are significantly larger than the nonadiabatic ones because $c_J=\beta b_J$ with $\beta\ll 1$. However, the spatially averaged $H_{a,x}$ remains negligible compared with the spatially averaged $H_{na,x}$, because the antisymmetric distribution of $H_{a,x}$ evident in Fig. 4(a) largely cancels.

\begin{figure}[!htbp]
  \centering
  \includegraphics[width=\columnwidth]{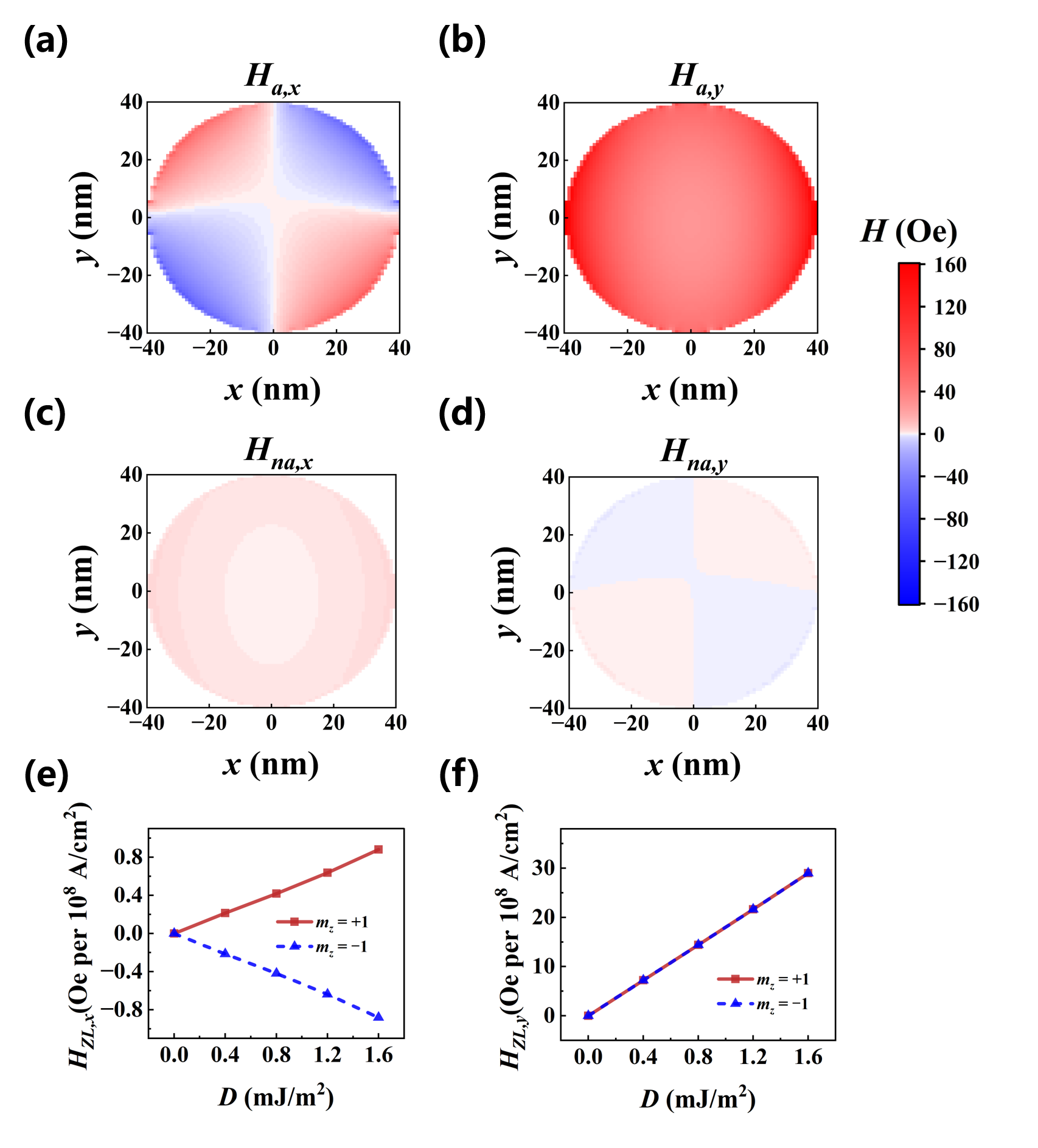}
  \caption{Effective fields of the Zhang--Li STT. (a), (b) Spatial distributions of the x- and y-components of the effective field arising from the adiabatic contribution of the Zhang--Li STT. (c), (d) Corresponding distributions for the nonadiabatic contribution. (e), (f) Effect of reversing m\textsubscript{z} on the x- and y-components of the total effective field for the Zhang--Li STT under different DMI strengths D.}
  \label{fig:4}
\end{figure}

Upon reversing the magnetization along the $z$-axis, $H_{\mathrm{ZL},x}$ changes sign while $H_{\mathrm{ZL},y}$ does not; increasing the DMI strength $D$ enhances the magnitudes of both components. The sign reversal of $H_{\mathrm{ZL},x}$ arises from the $\partial m_x/\partial x$ term in $H_{na,x}$, which changes sign under global magnetization reversal. The corresponding DMI energy density is given by Eq.~\eqref{eq:dmi-energy} \cite{r25}:
\begin{equation}
\resizebox{0.88\columnwidth}{!}{$\displaystyle
\varepsilon_{\mathrm{DM}}=D\left(m_z\frac{\partial m_x}{\partial x}-m_x\frac{\partial m_z}{\partial x}+m_z\frac{\partial m_y}{\partial y}-m_y\frac{\partial m_z}{\partial y}\right).
$}
\label{eq:dmi-energy}
\end{equation}
In contrast, $H_{\mathrm{ZL},y}$ remains nearly unchanged because it is dominated by the much larger adiabatic component $H_{a,y}$, which is invariant under global magnetization reversal.

Comparing the effective fields of the Zhang--Li STT with those of the SOT provides valuable insight into the anomalous thickness dependence of the SOT field usually observed in magnetic alloys \cite{r15,r45}. The SOT field consists of the damping-like component \ensuremath{\mathbf{H}_{\mathrm{DL}}\sim\frac{J}{t_{\mathrm M}}(\mathbf{m}\times\boldsymbol{\sigma})} and the field-like component \ensuremath{\mathbf{H}_{\mathrm{FL}}\sim\frac{J}{t_{\mathrm M}}\boldsymbol{\sigma}} \cite{r15,r44}. For a current applied along the x-axis, the SHE generates a spin current with polarization along the y-axis. Consequently, in a perpendicularly magnetized system under a weak external current, the \ensuremath{\mathbf{H}_{\mathrm{DL}}} and \ensuremath{\mathbf{H}_{\mathrm{FL}}} are oriented along the x- and y-directions, respectively. Reversing the magnetization reverses the sign of \ensuremath{\mathbf{H}_{\mathrm{DL}}} but leaves \ensuremath{\mathbf{H}_{\mathrm{FL}}} unchanged, consistent with the results shown in Figs. 4(e) and 4(f). In principle, both \ensuremath{\mathbf{H}_{\mathrm{DL}}} and \ensuremath{\mathbf{H}_{\mathrm{FL}}} scale inversely with the magnetic layer thickness t\textsubscript{M}. However, experiments on certain magnetic alloys have reported a proportional dependence on t\textsubscript{M} \cite{r15,r45}, a trend that remains not fully understood. Notably, an increase in t\textsubscript{M} is often accompanied by a growing composition gradient along the thickness direction of the alloy film, which can enhance the iDMI energy \cite{r23,r45}. This observation aligns with our findings in Figs. 4(e) and 4(f). Given that the magnetic alloy itself serves as the primary current channel, we suggest that the Zhang--Li STT arising from composition-graded iDMI may also play a significant role alongside SOT in such systems.

Finally, a brief discussion about application is in order. In addition to the simulation based on 0 K, we also introduced thermal fluctuation at 300 K in the simulation and compared it with SOT (see the Supplemental Material S6). Based on the reported experimental conditions of Pt/Co/AlO\textsubscript{x} \cite{r18}, our simulation results show that the J\textsubscript{c} for Zhang-Li STT and SOT are at the magnitudes of 10\textsuperscript{12} A/m\textsuperscript{2} and 10\textsuperscript{11} A/m\textsuperscript{2}, and the former is close to the experimental data in similar NM/M systems \cite{r9,r18,r46,r47,r48}. In the typical W/CoFeB/MgO system used in SOT-MRAM, however, the Zhang-Li STT is not the dominant mechanism for switching because the spin polarization in CoFeB is positive and the DW always moves against the electron flow \cite{r49,r50}, which is against STT behavior. Therefore, in W/CoFeB/MgO, SOT may be dominant over Zhang-Li STT due to the larger spin Hall angle in W and much weaker iDMI as compared to Pt/Co. Nevertheless, considering possible contribution from Zhang-Li STT may also be good for improving the crucial performance of the SOT-MRAM, such as the J\textsubscript{c} and switching time.

In conclusion, we have identified and theoretically validated the Zhang-Li torque as the driving force for deterministic current-driven magnetization switching in magnetic thin film with iDMI. By exploiting the iDMI-induced boundary DWs, this torque can produce all the essential signatures of SOT switching, including field-assisted reversal and the control of switching polarization, without invoking any spin current from the NM layer. Our findings establish that the Zhang-Li torque constitutes a third class of current-induced switching mechanisms, operating alongside Slonczewski STT and SOT. More broadly, our work also opens an avenue for device engineering---where the Zhang-Li torque can be either harnessed for low-power switching in iDMI-strong materials or mitigated in SOT-dominant designs. This insight offers a fresh and essential perspective for the physics and application of spintronic devices.

\section*{Acknowledgments}
This work was supported by the National Natural Science Foundation of China (Grant No. 12574119, 12304152), the open research fund of Songshan Lake Materials Laboratory (Grant No. 2023SLABFN26), the Natural Science Foundation of Hubei Province (No. JCZRYB202500910), and the Wuhan City Key R\&D Program (Grant No. 2025050602030069).

\end{document}